\documentclass[12pt,a4paper,english,superscriptaddress,aps,nofootinbib]{revtex4}
\usepackage[utf8]{inputenc}
\usepackage[T1]{fontenc}
\usepackage{amsmath,amssymb,graphicx}
\makeatletter
\usepackage{babel}
\usepackage[active]{srcltx}
\usepackage{graphicx,color}
\usepackage{changebar}
\usepackage{hyperref}
\usepackage[T1]{fontenc}
\usepackage{esint}
\usepackage{multirow}
\usepackage{dsfont}
\usepackage{ae}
\usepackage{amsmath}
\usepackage{braket}
\usepackage{mathtools}
\usepackage{slashed}
\usepackage{empheq}
\usepackage{multirow}
\usepackage{graphicx}
\usepackage{amsfonts}
\usepackage{amsmath}
\newcommand{\hodge}{{\star}}

\begin{document}
\title{Einstein-Bumblebee-Dilaton black hole solution }

\author{L. A. Lessa}
\email{E-mail: leandrolessa@fisica.ufc.br}
\affiliation{Universidade Federal do Cear\'{a} (UFC), Departamento de F\'{i}sica - Campus do Pici, Fortaleza, CE, C. P. 6030, 60455-760, Brazil.}

\author{J. E. G. Silva}
\email{E-mail: euclides@fisica.ufc.br}
\affiliation{Universidade Federal do Cear\'{a} (UFC), Departamento de F\'{i}sica - Campus do Pici, Fortaleza, CE, C. P. 6030, 60455-760, Brazil.}









\begin{abstract}

We obtain new black hole solutions in a Einstein-Bumblebee-scalar theory. By starting with a Einstein-Bumblebee theory in $D+d$ dimensions, the scalar dilaton field and its interaction with the gravitational and bumblebee fields are obtained by Kaluza-Klein (KK) reduction over the extra dimensions. Considering the effects of both the bumblebee vacuum expectation value (VEV) and the fluctuations over the VEV, we obtained new charged solutions in $(3+1)$ dimensions. For a vanishing dilaton, the black hole turned out to be a charged de Sitter-Reissner-Nordstrom solution,
where the transverse mode is the Maxwell field and the longitudinal mode is the cosmological constant. The stability of these new solutions is investigated by means of the analysis of the black hole thermodynamics. The temperature, entropy and heat capacity show that these modified black holes are thermodynamic stable. 

\end{abstract}

\maketitle


\section{Introduction}

The ongoing quest to identify potential violations of Lorentz symmetry remains a pivotal endeavor. The discovery of any departures from this symmetry would mark the advent of a new phase in physics. Some models in string theory \cite{corda}, very special relativity \cite{vsr}, noncommutative spacetime \cite{noncommutative} and loop quantum gravity \cite{lqg}, among others, enable Lorentz symmetry violation. The Standard Model Extension (SME) presents a structured framework for the exploration of theories involving Lorentz violations. Within the SME, Lorentz violation coefficients induce deviations from particle Lorentz symmetry \cite{sme}. Local Lorentz violation can arise through a mechanism involving the spontaneous breaking of symmetry due to self-interacting tensor fields. The vacuum expectation value (VEV) of these tensor fields leads to background tensor fields, which, when coupled to Standard Model (SM) fields, give rise to violations of local Lorentz symmetry at the particle level \cite{kostelecky,altschul,lessa}. This spontaneous Lorentz violation also ensures that the LV terms in the Lagrangian satisfy the Bianchi identities, a crucial property in the context of the gravitational field \cite{kostelecky}.

An instance within the gravitational sector of the SME that exemplifies the spontaneous violation of Lorentz symmetry is the Bumblebee field. 
This field, classified as a self-interacting vector field, possesses a vacuum expectation value (VEV) denoted as $b_{\mu}$, which establishes a preferred direction in spacetime \cite{ks}. We find in the literature numerous works that use this field. Within flat spacetimes, the causality and stability attributes of this model have been examined, encompassing both classical analyses \cite{ks2,bluhm,escobar} and investigations at the quantum level \cite{hernaski,maluf}. The spontaneous breaking of the Lorentz symmetry leads to the emergence of Nambu-Goldstone (NG) modes and massive modes \cite{ks2}. For a quadratic potential, in the so-called Kostelecky-Samuel (KS) model in $3+1$ dimensions, the fluctuations around the vev $b_{\mu}$ yield to two transverse NG modes and one longitudinal massive mode. Since only the transverse modes are propagating, the photon can be interpreted as a NG mode of the Bumblebee field instead of an elementary particle \cite{ks2,Seifert,seifert}.

In $3+1$ curved spacetimes, the effects of the Bumblebee non-minimally coupled to the gravitational field were studied for black holes \cite{bertolami1,casana,malufbh}, wormholes \cite{ovgun} and cosmology \cite{capelo}. In higher dimensions, the Bumblebee VEV modifies the Kaluza-Klein spectrum for bulk fields \cite{bertolami2,carroll,rizzo}. By assuming the Bumblebee field living in a AdS$_5$ bulk, the Ref. \cite{leandro} studied the effects of curvature of the bulk in a cosmological braneworld by modifying the dynamics of Bumblebee fluctuations. For a generalized Bumblebee dynamics, an analysis of the fluctuations was performed in Ref.\cite{sei}.

In this work, we are interested in study the effects of minimal coupling of the  scalar dilaton field to Bumblebee fluctuations for a black hole metric with static and spherical symmetry. This coupling can be motivated through a Kaluza-Klein reduction over the extra dimensions. By assuming a VEV in the radial direction, we obtain a new charged solution in $3 + 1$ dimensions which recovers the Sitter-Reissner-Nordstrom
solution when the dilaton field vanishes. In this scenario, the transverse Bumblebee mode assumes the role of the Maxwell field, while the longitudinal mode takes on the role of the cosmological constant. The connection between the cosmological constant and the massive mode establishes an upper limit for this mode, approximately $\beta_0^2 \sim 10^{-17} eV^{2}$. The novel solution exhibits geometric characteristics dependent on the value of the dilaton coupling constant. When $\tilde{a}<1/2$ , a black hole solution with an event horizon emerges, albeit being a singular solution at the origin. In investigating the stability of the solution, we examined thermodynamic parameters such as specific heat and isothermal charge susceptibility. Our analysis demonstrated that, when the Kaluza-Klein reduction on a circle occurs, the solution achieves thermodynamic stability while becoming electrically unstable. Moreover, our investigation confirms the validity of the first law of thermodynamics.

The work is organized as the following. In section \ref{sec1} we build the model that couples the dilaton field minimally to the Bumblebee fluctuations for a static and spherically symmetric spacetime. In section \ref{sec2}, we look for black hole solutions in $3+1$-dimensions for this model. In section \ref{sec3},we study the thermodynamics and stability of the black hole solution found. Final remarks are summarized in section \ref{sec4}. We will be using units where the speed of light, Planck’s constant, and Boltzmann’s constant equals unity, $c$ = $\hbar$ = $k$ = 1. Throughout the text, we adopt the capital Roman indices $(A,B,... = 0,1,2,3,4)$ denote 5-dimensional bulk spacetime indices, the Greek indices $(\mu, \nu, ... = 0, 1, 2, 3)$ the spacetime indices of the worldbrane. Moreover, we shall also take the Lorentzian signature for the spacetime metric to be $(-, +, +, +)$.
\section{Einstein-Bumblebee-Dilaton model}\label{sec1}

In this section we introduce the action for a modified Einstein-Bumblebee model including the dilatonic couplings.
Then, we obtain the action considering the terms steaming from the fluctuations of the Bumblebee on the vacuum expectation value $b_\mu$.

We start by defining the action for the Bumblebee field $B_\mu$ minimally coupled with gravity and coupled with the dilaton field $\phi$ in the form
\begin{align}\label{action1}\nonumber
    &S_{KS} = \int d^{4}x \sqrt{-g}\bigg[\frac{1}{2 \kappa^{2}}\bigg(R - \frac{1}{2}(\partial\phi)^2 \bigg) - \frac{e^{2\Tilde{a}\phi}}{4}B^{\mu\nu}B_{\mu\nu} \\
    &- \frac{\lambda e^{2\Tilde{a}\phi}}{2}(B^{\mu}B_{\mu}- b^2)^2 \bigg], 
\end{align}
where $\kappa^2=8\pi G$ with $G$ the Newton gravitational constant and $B_{\mu}$ is a vector field known as Bumblebee and $B_{\mu\nu}=\partial_{[\mu}B_{\nu]}$ is its field strength. The quadratic potential chosen induces the spontaneous Lorentz symmetry violation, where $\lambda$ is a mass dimension one positive self-interaction coupling constant, $b^{2}$ is a positive constant with squared mass dimension and the $\pm$ sign meaning if $b_{\mu}$ is spacelike or timelike. Moreover, the vacuum condition $V=0$ implies the existence of a vacuum expectation value (VEV) $<B_{\mu}>=b_{\mu}$ is the form
\begin{equation} \label{norma}
g^{\mu\nu}b_{\mu}b_{\nu} = \mp b^{2}.
\end{equation}
The key novelty in the action Eq.(\ref{action1}) is the presence of the massless scalar field $\phi$, the so-called dilaton. The dilaton coupling introduces a kind of damping into the Bumblebee dynamics. The dilaton interaction in the Einstein-Maxwell-scalar model has been extensively studied \cite{d1,d2,d3}. Here we extend these analysis to the Bumblebee field and its fluctuations. The action in Eq.(\ref{action1}) can be formally obtained by Kaluza-Klein dimensional reduction, as we performed in the appendix \ref{apen}.  It is worthwhile to mention that in the action Eq.(\ref{action1}) there is no non-minimal coupling of type $B^\mu B^\nu R_{\mu\nu}$. Therefore, the analysis and result obtained here differs significantly from Ref. \cite{casana}.

\subsection{Bumblebee fluctuations in a static and spherically symmetric spacetime}

From the action Eq.(\ref{action1}), we can see that if the Bumblebee field in on its VEV, i.e., for $B_\mu =b_\mu$, then the LV potential vanishes. Accordingly, no LV effects are present and we recover the Einstein-Maxwell-dilaton theory. However, considering small fluctuations around the VEV $b_\mu$, LV effects due to the Bumblebee fluctuations are expected to appear.

Let us consider the Bumblebee fluctuation around the vacuum expectation value (VEV) $b_\mu$ of form
\begin{equation} \label{flutuação}
B_{\mu} \approx b_{\mu} + \chi _{\mu},
\end{equation}
where $<B_{\mu}> = b_{\mu}$. Since the VEV defines a preferred direction in spacetime, we can decompose $\chi_{\mu}$ into transverse $A_{\mu}$ and longitudinal $\beta$ modes with respect to $b_{\mu}$ \cite{kos1} \begin{equation} \label{decomp}
\chi _{\mu} = \Tilde{A}_{\mu} + \beta \hat{b}_{\mu},
\end{equation}
where by defining the projection operators $P^{||}_{\mu\nu} = \frac{b_{\mu} b_{\nu}}{b^{\alpha}b_{\alpha}}$ and $P^{\perp}_{\mu\nu} = g_{\mu\nu} - \frac{b_{\mu} b_{\nu}}{b^{\alpha}b_{\alpha}}$,
we have $\Tilde{A}_{\mu} = P^{\perp}_{\mu\nu} \chi^{\nu}$ and $\beta\hat{b}_{\mu} = P^{||}_{\mu\nu} \chi^{\nu}$.
As result, we have to $\Tilde{A}_{\mu}b^{\mu}\approx0$ and $\hat{b}_{\mu}\hat{b}^{\mu} = \mp1$, where $\hat{b}_{\mu} = \frac{b_{\mu}}{\sqrt{b^{2}}}$.  Using the decomposition [\ref{decomp}], the smooth quadratic potential term becomes
\begin{equation}
V \approx 2 \lambda[(\hat{b} ^{\alpha}b_{\alpha})\beta]^{2}
\end{equation}
i.e., $V(X)\neq0$, therefore the $\beta$ is the massive mode.

As shown by Ref.\cite{leandro}, the longitudinal and transverse modes are highly coupled in general curved spacetimes. However, it is possible to simplify the analysis to spacetimes with some symmetries. Thus, let us assume a static and spherically symmetric spacetime of form 
\begin{equation}\label{metric}
ds^{2}  = -f(r) dt^2 + \frac{dr^2}{f(r)} + R(r)^2 d\Omega^2
\end{equation}
Assuming that the VEV has a preferred direction only in the radial direction, it is straightforward to show that the only non-vanishing component of VEV only depends on the radial coordinate, so we have that $b_{\mu\nu}=\partial_{[\mu}b_{\nu]}=0$. Moreover the $\Tilde{A}^r=0$, due to the decomposition $b_{\mu}\Tilde{A}^{\mu}=0$. Considering all of the above, we substitute the decomposition (\ref{decomp}) into action (\ref{action1}) and assume the spacetime (\ref{metric}). This leads us to the following linearized action with respect to Bumblebee fluctuations:
\begin{align} \label{action2}\nonumber
    &S_{KS} \approx  \int d^{4}x \sqrt{-g}\bigg[\frac{1}{2 \kappa^2} \bigg(R - \frac{1}{2}(\partial\phi(r))^2 \bigg) - \frac{e^{2 \Tilde{a}\phi(r)}}{4}\Tilde{F}_{\mu\nu}\Tilde{F}^{\mu\nu} \\
    &- 2\lambda b^2\beta_0^2 e^{2 \Tilde{a}\phi(r)} \bigg], 
\end{align}
where $\Tilde{F}_{\mu\nu}(r)=\partial_{[\mu}\Tilde{A}_{\nu]}$ with $\Tilde{A}^r=0$.

It is worthwhile to mention that there is no kinetic term for the longitudinal mode $\beta$. This is precisely due to the choice of VEV and the static and spherical symmetry of spacetime \cite{leandro}. Therefore, the $\beta=\beta_0$ is a constant in this configuration. This result agrees with the fact that, in flat spacetimes, the longitudinal mode $\beta$ is non-propagating.

The action in Eq.(\ref{action2}) resembles the Einstein-Maxwell-dilaton action with a cosmological constant term $\Lambda=\lambda b^2 \beta_{0}^{2}$. This cosmological constant term is also damped by the dilaton interaction. Therefore, the transverse Bumblebee mode $\tilde{A}_\mu$ is similar to the gauge field and the longitudinal mode
$\tilde{\beta}$ provides a source for a cosmological constant. Through this identification, we can compute an upper bound for the non-dynamic massive mode $\beta_0$ in $D=4$. By combining the upper limit established by Ref. \cite{casana} in the classical test of Time delay of light for the parameter $ \lambda b^2 \sim 10^{-19}$ with the experimental assessment of the cosmological constant $c^2\Lambda \sim 10^{-36} eV^{2}$, as estimated by \cite{planck}, we conclude that the magnitude of the massive mode is bounded from above by an order of $\beta_0^2 \sim 10^{-17} eV^{2} $.

\section{(3+1)-dimensional black hole solution} \label{sec2}

In this section, we seek for black hole solutions in (3+1)-dimensions for an Einstein-Bumblebee-Dilaton spacetime described by the metric action (\ref{action2}). 
By varying this action with respect to the metric, the vector field, and the scalar field, we obtain the following equations, respectively:
\begin{align} \label{e1}\nonumber
&\Tilde{G}_{\mu\nu} \equiv G_{\mu\nu} - \kappa^2 e^{2\Tilde{a}\phi} \bigg( \Tilde{F}_{\mu} \ ^{\alpha}\Tilde{F}_{\nu\alpha} - \frac{1}{4}\Tilde{F}_{\lambda\sigma}\Tilde{F}^{\lambda\sigma}g_{\mu\nu} - V_0 g_{\mu\nu} \bigg) \\
&- \frac{1}{2}D_{\mu} \phi D_{\nu}\phi + \frac{1}{4} D_{\sigma}\phi D^{\sigma}\phi g_{\mu\nu}=0,
\end{align}
\begin{equation}\label{e2}
    D_{\mu}(e^{2\Tilde{a}\phi}\Tilde{F}^{\mu\nu})=0,
\end{equation}
\begin{equation}\label{e3}
    D_{\sigma}D^{\sigma}\phi = 4 \kappa^2 \Tilde{a}e^{2\Tilde{a}\phi}\bigg(\frac{1}{4}\Tilde{F}_{\lambda\sigma}\Tilde{F}^{\lambda\sigma}+V_0 \bigg),
\end{equation}
where the LV potential term $V_0$ is given by 
\begin{equation}
    V_0 \equiv 2\lambda b^2\beta_0^2.
\end{equation}
Note that Lorentz violation (LV) results in a positive potential, as it is necessary for $\lambda > 0$ to induce a spontaneous breaking of Lorentz symmetry.

From the action Eq.(\ref{action2}), the transverse mode has a similar dynamics to the gauge field.
Consequently, we can consider the existence of an isolated electric charge described by Eq. (\ref{e2}), and the corresponding field strength has the form
\begin{equation} \label{elet}
    \Tilde{F}_{tr}=\frac{\Tilde{q} e^{-2\Tilde{a}\phi}}{R(r)^2} ,
\end{equation}
where $\Tilde{q}$ is a constant of integration that relates to an analogue of electric charge through the formula $\Tilde{Q}=\frac{1}{4\pi}  \int_{S^2} e^{2\Tilde{a}\phi} \hodge \Tilde{F}$, where $\hodge$ is the Hodge dual and the $S^2$  is any two-dimensional sphere.

The equation of the motion (EOM) (\ref{e1}), (\ref{e2}), and (\ref{e3}) are typically not integrable. To overcome this issue, we adopt the same approach as done in Ref. \cite{k}, where the function $R$ is given by
\begin{equation}\label{r}
    R(r) = r ^ N,
\end{equation}
where $N$ is a constant. As a result, we end up with four independent equations\footnote{Three independent equations from (\ref{e1}) and one additional equation from (\ref{e3}), considering that Eq. (\ref{e2}) has already been solved using the solution (\ref{elet}).} and three variables that need to be determined: $\phi$, $N$, and $f$. Therefore, our system is potentially integrable. We can express the equations of motion as:
\begin{align}\label{eq1} \nonumber
    &\Tilde{G}^{t} \ _t = \frac{1}{ r ^ 2N} - \frac{N(3N-2)f}{r^2} - \frac{N f'}{r} \\
    &-  \kappa^2  e^{-2\Tilde{a}\phi}\bigg( \frac{\Tilde{Q}^2}{2  r ^{4N}} + V_0 e^{4\Tilde{a}\phi}\bigg) - \frac{1}{4} f (\phi ')^2=0,
\end{align}
\begin{align}\label{eq2}\nonumber
    &\Tilde{G}^{r} \ _r = \frac{1}{r ^ 2N} - \frac{N^{2} f}{r^2} - \frac{N f'}{r} \\
    &-  \kappa^2  e^{-2\Tilde{a}\phi}\bigg( \frac{\Tilde{Q}^2}{2  r ^{4N}} + V_0 e^{4\Tilde{a}\phi}\bigg) + \frac{1}{4} f (\phi ')^2=0,
\end{align}
\begin{align}\label{eq3}\nonumber
    & \Tilde{G}^{\theta} \ _{\theta}= \frac{2N(N-1)f}{2 r^2} + \frac{N f'}{r} + \frac{f''}{2} \\
    &-  \kappa^2  e^{-2\Tilde{a}\phi}\bigg( \frac{\Tilde{Q}^2}{2  r ^{4N}} - V_0 e^{4\Tilde{a}\phi}\bigg) + \frac{1}{4} f (\phi ')^2=0,
\end{align}
\begin{equation}\label{eq4}
\frac{2N f \phi ' }{r} + \phi 'f ' + f \phi'' = 4 \kappa^2 \Tilde{a}e^{2\Tilde{a}\phi} \bigg( - \frac{\Tilde{Q}^2}{2  r ^{4N}} e^{-4\Tilde{a}\phi} + V_0 \bigg),
\end{equation}
where a prime stands for the derivative with respect to $r$.

\subsection{Decoupling limit}
An interesting case arises when the Bumblebee fluctuations decouple from the dilaton field, i.e., when $\tilde{a}=0$. In the action (\ref{action2}), we recover an Einstein-Maxwell term with a massless scalar and a cosmological constant term $\Lambda=\lambda b^2\beta_0^2$. Thus, we expect to obtain a De Sitter-Reissner-Nordstron solution. 

Indeed, from Eq.(\ref{eq1}) and Eq.(\ref{eq2}) with $  \Tilde{G}^{t} \ _t -  \Tilde{G}^{r} \ _r $ and assuming the Eq. (\ref{r}), we obtain the dilaton of form
\begin{equation}\label{escalar}
    \phi(r) = \phi_{0} + 2 \sqrt{N(1-N)}\text{ln} r,
    \end{equation}
where $\phi_{0}$ is a constant. Substituting the dilaton solution Eq.(\ref{escalar}) into Eq.(\ref{eq3}) and Eq.(\ref{eq4}), we obtain the line element
\begin{align} \label{sol1}\nonumber
  &  ds^2 = -\bigg( 1 - \frac{r_S}{r} + \frac{Q}{r^2} - \frac{\lambda b^2 \beta_0^2}{3}r^2 \bigg) dt^2 \\
    &+ \bigg(  1 - \frac{r_S}{r} + \frac{Q}{r^2} - \frac{\lambda b^2 \beta_0^2}{3}r^2  \bigg) ^{-1}dr^2 + r^2 d\Omega^2_{2}.
\end{align}
From Eq.(\ref{escalar}), the condition $N=1$ leads to a constant dilaton. Accordingly, the Einstein-Maxwell-De Sitter action is recovered from the action in Eq.(\ref{action2}).

It is worthwhile to mention that the Bumblebee transverse mode fluctuation $A_\mu$ give rises to an electrically charged black hole, whereas the massive longitudinal mode is a source for a cosmological constant term. In flat spacetime, the transverse Bumblebee mode is a massless field which can be identified with the photon. On the other hand, the longitudinal mode $\beta$ is a massive field which can be shown to be non propagating. The Eq.(\ref{sol1}) reveals that, in the strong field regime, the transverse Bumblebee mode can still be identified with the photon, for it provides the same metric solution. However, the longitudinal massive mode acquires a new interpretation. Despite being non-propagating, this mode provides a source for the cosmological constant.

\subsection{Dilatonic solutions}

Once studied the solutions including the Lorentz violating terms only, let us consider the influence of the dilaton scalar field steaming from the extra dimensions.

By substituting the equation (\ref{escalar}) into equations (\ref{eq1}) to (\ref{eq4}), we obtain the following solutions:
    \begin{equation}\label{s1}
        N= \frac{4 \Tilde{a}^2}{1+4 \Tilde{a}^2},
        \end{equation}
\begin{equation}\label{s2}
    f(r) = r^{\frac{1-4\Tilde{a}^2}{1+4\Tilde{a}^2}}\bigg[ - \frac{(1 + 4\Tilde{a}^2)  \mathcal{M}}{4\Tilde{a}^2}  + \bigg(\frac{1+4\Tilde{a}^2}{1-4\Tilde{a}^2}\bigg)  (\Tilde{Q}^2\kappa^2 e^{2\Tilde{a}\phi_0}-1)r\bigg],
\end{equation}
\begin{equation}\label{s3}
       V_0 \equiv 2\lambda b^2\beta_0^2 =\frac{\bigg(   2e^{2\Tilde{a}\phi_0}-\Tilde{Q}^2\kappa^2(1+4\Tilde{a}^2) \bigg)}{2 (1-4\Tilde{a}^2)\kappa^2} e^{-4\Tilde{a}\phi_0},
\end{equation}
\begin{equation}\label{s4}
    \phi(r) = \phi_0 - \frac{4\Tilde{a}}{1+4\Tilde{a}^2}ln r,
\end{equation}
where $\mathcal{M}$ is an integration constant which is related to mass.  Using the quasilocal mass formalism \cite{m1,m2} with $\mathcal{M}$ as the reference background, we can calculate the mass $M$ of the solution as $\mathcal{M}=r_{S}$, where $r_{S}$ is the Schwarzschild radius, i.e., $r_{S}=2 M G$.

It is important to note that the metric reduces to the Schwarzschild black hole when $\Tilde{a}^2\rightarrow-\infty$.
Indeed, by Eq.(\ref{s1}), $N=1$, i.e., we recover the usual spherical metric element. Moreover, Eq.(\ref{s4}) shows that the dilaton field is constant and thus, the kinetic term in the action Eq.(\ref{action2}) vanishes. Furthermore, the $f(r)$ metric function becomes
\begin{equation}
    f(r)\rightarrow 1- \frac{\mathcal{M}}{r},
\end{equation}
as $\tilde{a}\rightarrow -\infty$. Interestingly, the dilaton coupling $e^{\tilde{a}\phi}$ damped both the Lorentz-violating and the dilatonic effects in the regime $\tilde{a}\rightarrow -\infty$. In this overdamped limit, the charge and the cosmological constant effects steaming from the Bumblebee fluctuations are exponentially suppressed.

\begin{figure}[h] 
\includegraphics[height=6cm]{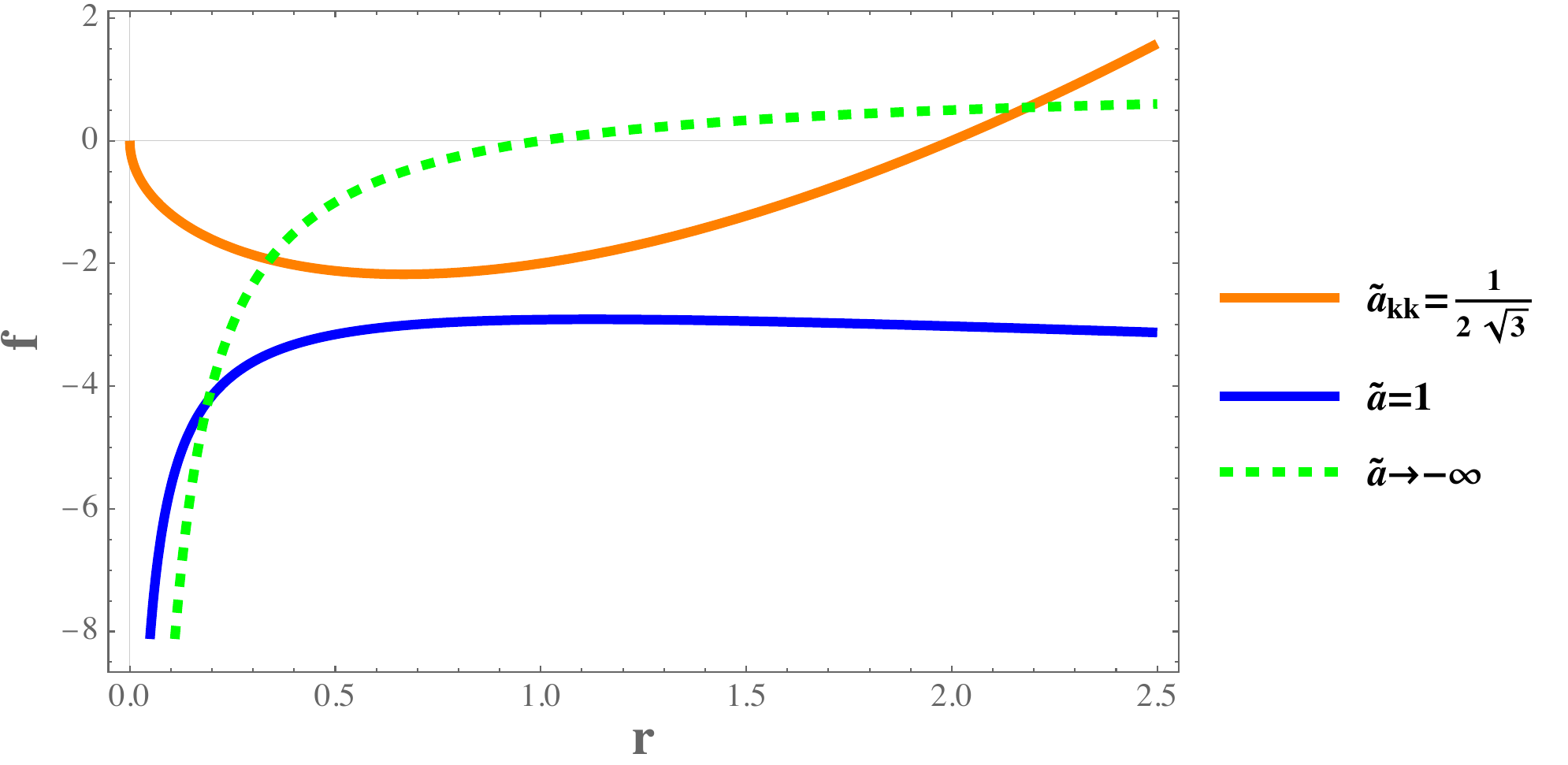}\\
           \caption{Plot of $f$ vs $r$ assuming that $\tilde{Q}^2=2$ and $\kappa^2=1$. The green dashed line is the Schwarzschild solution. We assume that $r_S=1$.} 
       \label{figsol}
\end{figure}

By performing a transformation $\Tilde{r} \rightarrow r ^{\frac{4 \Tilde{a}^2}{1+4 \Tilde{a}^2}} $ in Eq.(\ref{s2}), we obtain the following solution
\begin{align} \label{sol}\nonumber
    &ds^2 = -\bigg[ - \frac{(1 + 4\Tilde{a}^2) r_{S}}{4\Tilde{a}^2\Tilde{r}^{\frac{4\Tilde{a}^2-1}{4\Tilde{a}^2}}}+\bigg(\frac{1+4\Tilde{a}^2}{1-4\Tilde{a}^2}\bigg)  (\Tilde{Q}^2\kappa^2 e^{2\Tilde{a}\phi_0}-1)\Tilde{r}^{\frac{1}{2\Tilde{a}^2}} \bigg] dt^2 \\ \nonumber
    &+ \bigg(\frac{1+4\Tilde{a}^2}{4\Tilde{a}^2}\bigg)^2 \Tilde{r}^{\frac{1}{2\Tilde{a}^2}} \bigg[ - \frac{(1 + 4\Tilde{a}^2) r_{S}}{4\Tilde{a}^2\Tilde{r}^{\frac{4\Tilde{a}^2-1}{4\Tilde{a}^2}}}\\
    &+\bigg(\frac{1+4\Tilde{a}^2}{1-4\Tilde{a}^2}\bigg)  (\Tilde{Q}^2\kappa^2 e^{2\Tilde{a}\phi_0}-1)\Tilde{r}^{\frac{1}{2\Tilde{a}^2}} \bigg] ^{-1}d\Tilde{r}^2 + \Tilde{r}^2 d\Omega^2_{2}.
\end{align}
The metric function $f(r)$ given in Eq.(\ref{sol}) is quite cumbersome and thus, we plotted this function in the figure (\ref{figsol}) to study its behaviour. For $\Tilde{a}\rightarrow-\infty$ (green line), we obtain the usual Schwarzchild solution. For $\tilde{a}_{kk}=\frac{1}{2\sqrt{3}}$ (orange line), the function $f(r)$ is finite at the origin and it diverges as $r\rightarrow\infty$. As elucidated in the Appendix \ref{apen}, this particular value for the dilaton coupling constant finds its motivation in a Kaluza-Klein reduction. Thus, we have assigned a distinctive nomenclature to it.  Note that the black hole for this solution has only one horizon. On the other hand, for $\tilde{a}=1$ (blue line), the black hole has no horizon, i.e., $f(r_h )=0$.

\begin{figure}[h] 
 \includegraphics[height=6cm]{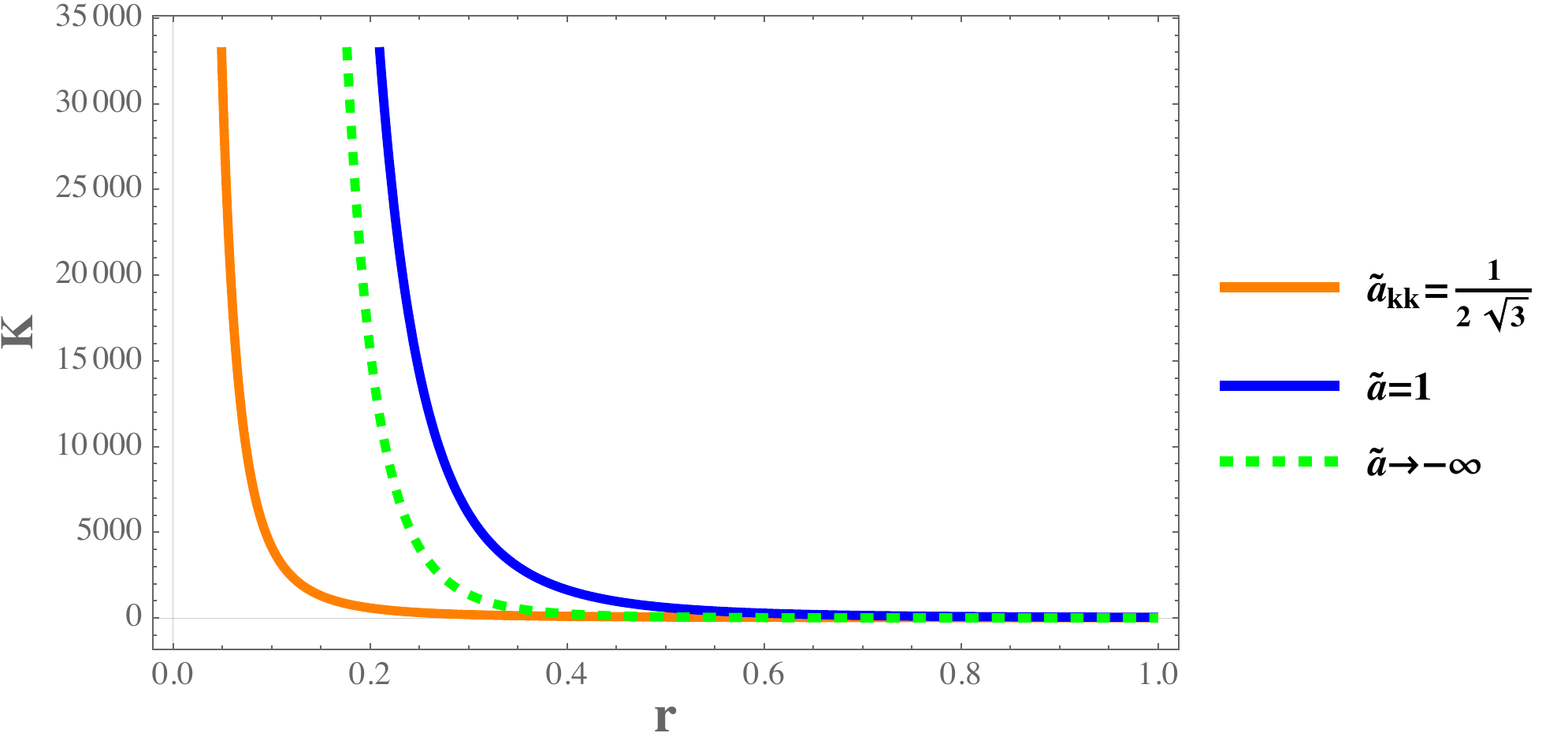}\\
           \caption{Plot of $K$ vs $r$ assuming that $\tilde{Q}^2=2$ and $\kappa^2=r_S=1$.} 
       \label{figk}
\end{figure}

In order to probe the singularities of the black hole solution given by Eq.(\ref{sol}), let us study the behaviour of the Kretchmann scalar $K=R_{\mu\nu\alpha\beta}R^{\mu\nu\alpha\beta}$ plotted in the figure (\ref{figk}). Note that the curvature diverges at the origin for all the values considered. Thus, the solution for $\tilde{a}=1$ (blue line) represents a naked singularity. Asymptotically, the curvature vanishes for the three cases. Thus, the dilaton coupling suppress the asymptotic effects of the cosmological constant term $\lambda b^2 \beta_{0}^2$ steaming from the Bumblebee longitudinal mode.

From Eq.(\ref{s2}) and Eq.(\ref{s3}), the dilaton parameter $\Tilde{a}$ must satisfy $\Tilde{a}\neq \frac{1}{2}$. Moreover, since $\lambda$ must be positive to induce spontaneous Lorentz symmetry breaking \cite{kos1}, the potential $V_0$ should also be positive. Thus, for 
$\tilde{a}<\frac{1}{2}$ we obtain the constrain
\begin{equation}\label{vin}
 \Tilde{Q}^2 \kappa^2    e ^{2\Tilde{a}\phi_0} \leq \frac{2}{1+4\Tilde{a}^2}.
\end{equation}
Conversely, for $\tilde{a} > \frac{1}{2}$, we have
\begin{equation}\label{vinc}
\tilde{Q}^2 \kappa^2 e^{2\tilde{a}\phi_0} \geq \frac{2}{1+4\tilde{a}^2}.
\end{equation}
Therefore, the Bumblebee charge $\tilde{Q}$ should be inversely proportional to the gravitational constant $\kappa$.

In order to obtain the event horizons for solution (\ref{sol}) just set that $g^{rr}=0$. Thus , we find only one horizon given by
\begin{equation}\label{event}
    r_h = \Tilde{r}_h ^{\frac{1+4\Tilde{a}^2}{4\Tilde{a}^2}}=  \bigg(\frac{1-4\Tilde{a}^2}{4\Tilde{a}^2} \bigg)\frac{r_{S}}{(\Tilde{Q}^2\kappa ^2 e^{2\Tilde{a}\phi_0}-1)}.
\end{equation}
Note that the constraint (\ref{vin}) always ensures that the horizon generated by the effects of the LV together with the dilaton is greater than the Schwarzschild horizon. 
As we can see in Fig.\ref{figsol}.

The solution for $\tilde{a}_{kk}=\frac{1}{2\sqrt{3}}$ has interesting properties, as shown in the figures (\ref{figsol}) and (\ref{figk}).
Indeed, this value can be obtained by dimensional reduction with one extra dimension compactified using the formalism developed by Scherk and Schwarz \cite{ss}. The line element for this configuration reads
\begin{align} \label{sol1}
    &ds^2_{kk} = -\bigg[ - 4 r_{S} \Tilde{r}^{2}+ 2 (8\pi l_{Pl}^2 \Tilde{Q}^2_{kk} e^{\frac{\phi_0}{\sqrt{3}}}-1)\Tilde{r}^{6} \bigg] dt^2 \\ \nonumber
    &+ 16 \Tilde{r}^{6} \bigg[ -4 r_{S} \Tilde{r}^{2}+ 2 (8\pi l_{Pl}^2\Tilde{Q}^2_{kk} e^{\frac{\phi_0}{\sqrt{3}}}-1)\Tilde{r}^{6} \bigg] ^{-1}d\Tilde{r}^2 + \Tilde{r}^2 d\Omega^2_{2},
\end{align}
whereas the relation between the Bumblebee charge and the LV VEV is given by
\begin{equation}\label{s32}
     V_0 \equiv 2\lambda b^2\beta_0^2 = \bigg( \frac{3 }{2}e^{\frac{\phi_{0}}{\sqrt{3}}} -8\pi l_{Pl}^2\Tilde{Q}^2_{kk}   \bigg)\frac{e^{\frac{-2\phi_{0}}{\sqrt{3}}}}{\kappa^2}.
\end{equation}
Here, we use the definition of the Planck length $l_{Pl}^2 = \frac{\kappa^2}{8\pi}$ and define the KK charge as
\begin{equation}
\tilde{Q}_{kk}^2 \equiv 2\pi R_y \tilde{Q}^2.
\end{equation}
where $R_y$ is the radius of extra dimension. Furthermore, assuming the above solutions, we find that the event horizon and the constraint to ensure spontaneous Lorentz symmetry breaking are given by
\begin{equation}\label{event1}
r_h = \tilde{r}_h^4 = \frac{2r_S}{(8\pi l_{Pl}^2 \tilde{Q}_{kk}^2 e^{\frac{\phi_0}{\sqrt{3}}}-1)},
\end{equation}
and
\begin{equation}\label{ine}
8\pi l_{Pl}^2 \tilde{Q}_{kk}^2 e^{\frac{\phi_0}{\sqrt{3}}}-1 \leq \frac{1}{2}.
\end{equation}
It is important to note that at the upper bound of the above inequality, the horizon $r_h$ (\ref{event1}) leads to a shift in the Schwarzschild radius, i.e., $r_h = 4r_S$.

Finally, let us examine how the solution (\ref{s2}) affects the radial component of the vacuum expectation value (VEV). By substituting Eqs. (\ref{escalar}) and (\ref{s2}) into Eq. (\ref{norma}), we find that
\begin{align}\nonumber
&b_{r} = b \bigg[ - \frac{(1 + 4\Tilde{a}^2) r_S}{4\Tilde{a}^2} r^{\frac{1-4 \Tilde{a}^2}{1+4 \Tilde{a}^2}}  \\
&+ \bigg(\frac{1+4\Tilde{a}^2}{1-4\Tilde{a}^2}\bigg)  (\Tilde{Q}^2\kappa^2 e^{2\Tilde{a}\phi_0}-1)r^{\frac{2}{4 a^2+1}}\bigg]^{-\frac{1}{2}}.
\end{align}
For $\tilde{a}<\frac{1}{2}$, the VEV component above vanishes as $r\rightarrow \infty$ and it diverges at the origin. And the opposite behavior happens with the case where $\Tilde{a}>\frac{1}{2}$.




\section{Thermodynamics} \label{sec3}
In this section, we calculate several important thermodynamic quantities to study the thermodynamic stability of the solution (\ref{sol1}). We focus on the KK dimensional reduction on a circle $S^1$ in our analysis, but the generalization to the solution (\ref{sol}) is straightforward. 
As shown by Ref.\cite{term1} , it is possible to associate thermodynamic stability with microscopic fluctuations of the system.  The stability conditions can be expressed as
\begin{equation} \label{estab}
    C_{\Tilde{Q}} \equiv T \bigg(\frac{\partial S}{\partial T}\bigg)_{\Tilde{Q}} \geq 0  \ , \ \chi_{T}  \equiv  \bigg(\frac{\partial \Tilde{Q}}{\partial \Tilde{\psi}}\bigg)_{T} \geq 0.
\end{equation}
The first quantity, $C_{\tilde{Q}}$, represents the specific heat at constant electric charge. It is an analogue of the specific heat at constant volume in fluid systems. The other quantity, $\chi_T$, is the isothermal charge susceptibility. We will explain and calculate each of these quantities later on.

\begin{figure}[h] 
 \includegraphics[height=6cm]{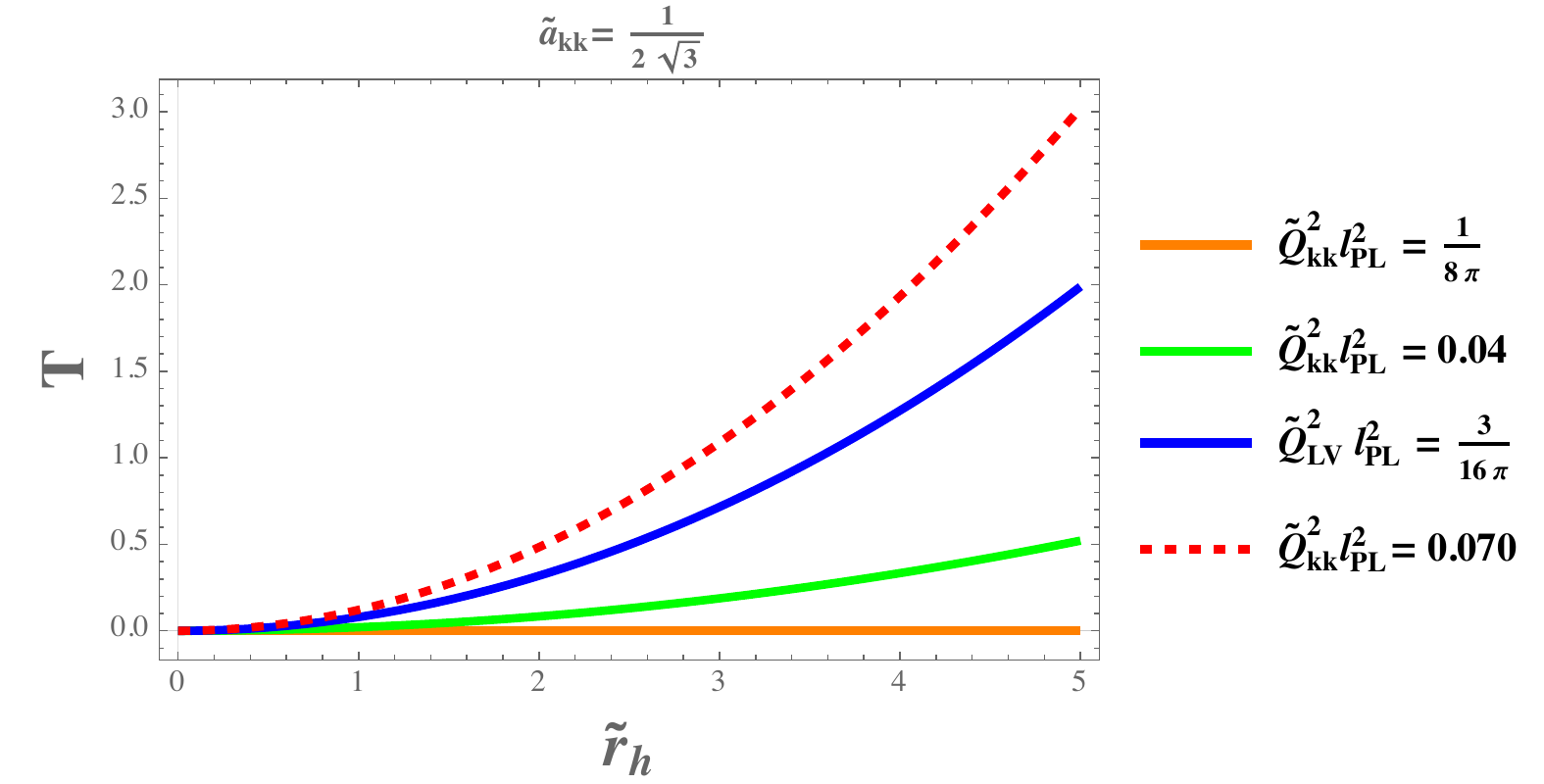}\\
           \caption{Plot of $T$ vs $\Tilde{r}_h$ assuming that  $\Tilde{a}_{kk}= \frac{1}{2\sqrt{3}}$.} 
       \label{fig1}
\end{figure}

Now, let us calculate the temperature of the black hole using the event horizon (\ref{event1}). However, before doing so, we can assume without loss of generality that $\phi_0=0$ from now on, as this constant is only an overall factor in the thermodynamic analysis. Therefore, the temperature can be calculated using the expression $T= \frac{f'(\tilde{r}_h)}{16\pi \tilde{r}_h^3}$ \footnote{We can calculate surface gravity $\kappa$, hence temperature $T= \frac{\kappa}{2\pi}$, for static spherically symmetric spacetime with metric given by
\begin{equation}
    ds^2 =  g_{tt}dt^2 + g_{rr}dr^2 + r^2d\Omega^2_{2},
\end{equation}
through the following formula
\begin{equation}
    \kappa = -\frac{\partial_r g_{tt}}{2\sqrt{-g_{rr} g_{tt}}}
\end{equation}}. Substituting the Eqs.(\ref{event1}) and (\ref{sol1}) in this expression, we obtain that
\begin{equation}\label{temp}
    T = \frac{(8\pi l_{Pl}^2\Tilde{Q}_{kk}^2-1)}{2\pi}\Tilde{r}_{h}^2.
\end{equation}
Thus, the temperature of the black hole decreases as its mass decreases. In Figure (\ref{fig1}), we plot the temperature as a function of the horizon for $\tilde{a}_{kk}=\frac{1}{2\sqrt{3}}$. It is evident that the temperature decreases as the horizon, and consequently the mass, decreases. Additionally, we ascertain that the extremal limit, characterized by a vanishing temperature, is attained when either $r_S=0$ or $8\pi l_{Pl}^2\tilde{Q}_{kk}^2=1$, in accordance with Eq. (\ref{event1}). Therefore, in the latter case, if the KK charge is on the order of the Planck mass, i.e., $\tilde{Q}_{kk}\sim M_{Pl}$, the black hole temperature becomes zero.

We can define a LV critical charge $\tilde{Q}_{LV}$, which indicates the effects of the spontaneous breaking of Lorentz symmetry. This critical charge is obtained by considering the upper bound of the inequality (\ref{ine}). When $\tilde{Q}_{LV}<\tilde{Q}_{kk}$, i.e., when $\lambda<0$ according to the constraint (\ref{vin}), the effects of LV are suppressed. In Figure (\ref{fig1}), this corresponds to the dashed line. For $\phi_0=0$, we can express the LV critical charge as:
\begin{equation}\label{lv1}
    \Tilde{Q}_{LV}^2\equiv\frac{3}{32\pi^2 R_y l_{Pl}^2}.
\end{equation}
Besides we can determine a LV critical temperature from Eq. (\ref{temp}). This temperature is given by
\begin{equation}\label{lv2}
    T_{LV} \equiv \frac{r_{S}^{1/2}}{2\pi}.
\end{equation}
Immediately, we can observe that the temperature (\ref{lv2}) is quite different from the temperature associated with the standard Schwarzschild black hole, which decreases with $r_S$.

Another crucial thermodynamic quantity related to the event horizon is the entropy. It can be expressed as $S = \frac{2\pi A}{\kappa^2}$, where $A$ is the area of the horizon. It is important to note that the entropy is determined by the volume of the 2-dimensional constant-time slices of the event horizon. Assuming that the horizon has a topology of $S^2$, the entropy for (\ref{event1}) is given by 
\begin{equation}
    S = \frac{\pi}{ l_{Pl}^2}\bigg( \frac{2 r_S}{(8\pi l_{Pl}^2\Tilde{Q}_{kk}^2-1)} \bigg)^\frac{1}{2}.
\end{equation}
Again, we can associate the entropy with the LV critical limit in the following way
\begin{equation}
    S_{LV} = \frac{2\pi}{ l_{Pl}^2} r_S ^\frac{1}{2}.
\end{equation}

It is important to highlight that the electric flux given by Eq. (\ref{elet}) is constant, as implied by Eqs. (\ref{s1}) and (\ref{s4}). As a result, the presented solution differs from the standard Reissner-Nordström (RN) solutions, where the electric charge term in the metric coefficient is proportional to $r^{-2}$. Additionally, the scalar potential associated with the horizon is defined using the following standard relation \cite{kill}:
\begin{equation}
    \Tilde{\psi}(r) = \Tilde{A}_{\mu}\chi^{\mu}\bigg|_{reference}-\Tilde{A}_{\mu}\chi^{\mu}\bigg|_{r \rightarrow r _h},
\end{equation}
where $\chi = C \partial_t$ is the null generator of the horizon, and $C$ is an arbitrary constant that 
it will be fixed next. By substituting Eq. (\ref{elet}) into the above equation, we find that the scalar potential grows linearly with radius, i.e., $\tilde{\psi}(r) = A\tilde{Q}e^{-2\tilde{a}\phi_0}r + \tilde{\psi}_0$, where $\tilde{\psi}_0$ is a constant given by:
\begin{equation}\label{potencial}
  \Tilde{  \psi}_0 = C \Tilde{Q}e^{-2\Tilde{a}\phi_0} r_h.
\end{equation}

Assuming the grand canonical ensemble, where the potential $\tilde{\psi}$ is fixed at the boundary with a value of $\tilde{\psi}_0 = C\tilde{Q}\tilde{r}_h^4$, and considering the energy $E$ in this ensemble to be the black hole mass $M$, it can be verified that the first law of thermodynamics holds,
\begin{equation}
    dM =  T dS + \Tilde{\psi}_0d\Tilde{Q},
\end{equation}
if we fix the constant in Eq.(\ref{potencial}) as $C=4\pi \sqrt{2\pi R_y}$.

\begin{figure}[h] 
 \includegraphics[height=6cm]{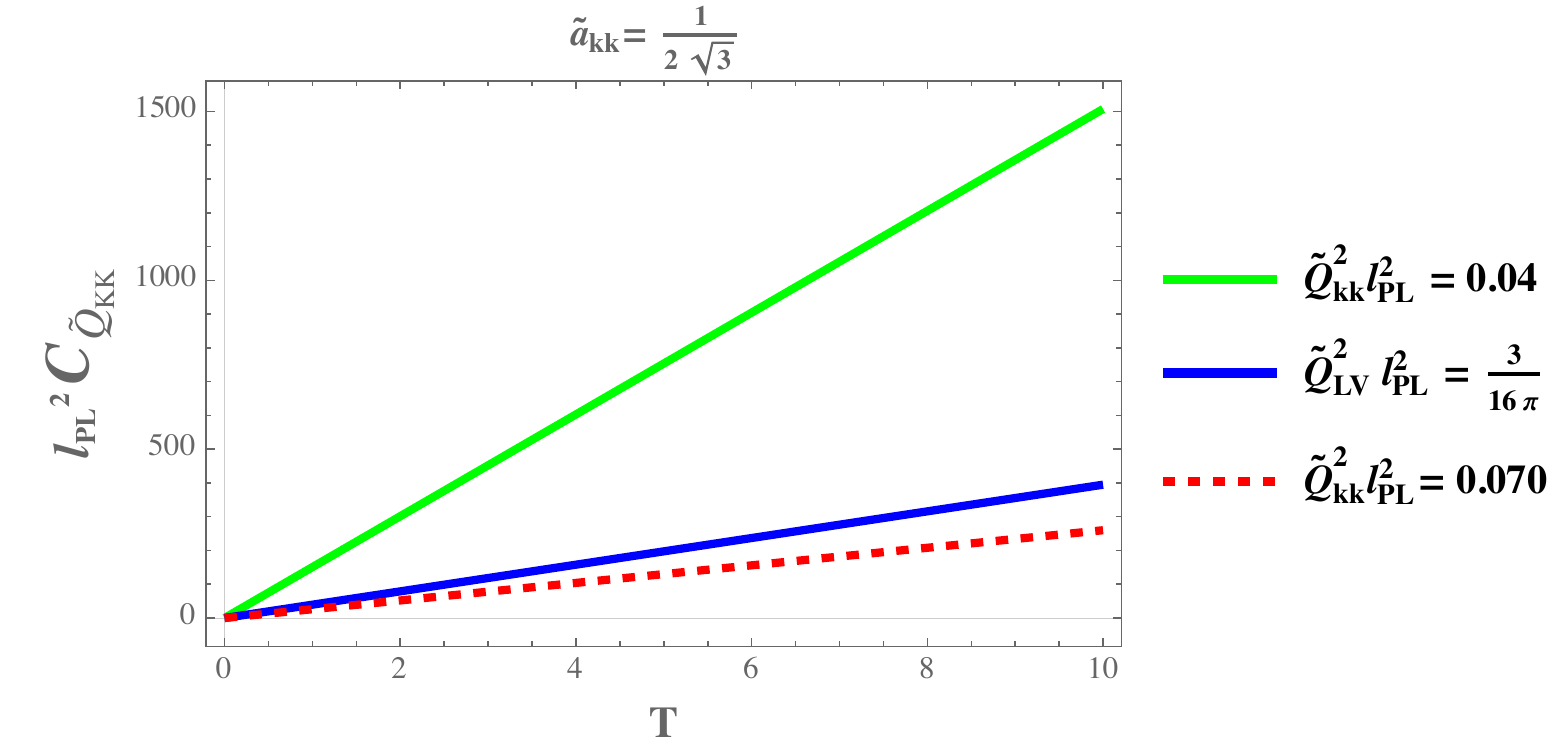}\\
           \caption{Plot of $l_{PL}^2 C_{\Tilde{Q}_{kk}}$ vs $T$ assuming that $\Tilde{a}_{kk}=\frac{1}{2\sqrt{3}}$.} 
       \label{fig4}
\end{figure}

Now we can study the stability of the solution (\ref{sol1}). First we consider the canonical ensemble, whose charge $\Tilde{Q}_{kk}$ is held fixed. 
For this ensemble, we can define the heat capacity at constant charge as $    C_{\Tilde{Q}_{kk}}= T (\frac{\partial S}{\partial T} )_{\Tilde{Q}_{kk}}$. Considering the temperature (\ref{temp}), the heat capacity is given by
\begin{equation}\label{cp}
    C_{\Tilde{Q}_{kk}} =\frac{\pi}{l_{Pl}^2} \bigg(\frac{2r_S}{(8\pi l_{Pl}^2\Tilde{Q}_{kk}^2 -1)}\bigg)^\frac{1}{2}.
\end{equation}
For a better understanding of this quantity, we can rewrite it as a function of temperature (\ref{temp}), instead of the horizon. Substituting the Eq.(\ref{temp}) in Eq.(\ref{cp}), we get that
\begin{equation}
     C_{\Tilde{Q}_{kk}} = \frac{2 \pi ^2 T}{l_{Pl}^2(8\pi l_{Pl}^2\Tilde{Q}_{kk}^2 -1)}.
\end{equation}
Thus, the thermodynamic stability is always achieved when $8\pi l_{Pl}^2\Tilde{Q}_{kk}^2>1$, as is shown in Fig. \ref{fig4}. 
More, the LV critical heat capacity is given by
\begin{equation}
    C_{\Tilde{Q}_{LV}} =\frac{ 2 \pi }{ l_ {Pl}^2} r_S ^\frac{1}{2}.
\end{equation}
The curve associated with this quantity is the thin blue line in Fig. \ref{fig4}.

Another important condition that determines the electrical stability of the black hole, i.e., whether the solution is stable against electrical fluctuations, is given by the isothermal charge susceptibility $\chi_T$ \cite{h1}. To analyze this condition, we need to obtain the isothermal curves of the equation of state $T = T(\tilde{\psi}, \tilde{Q})$. The first step is to express the temperature (\ref{temp}) as a function of the potential $\tilde{\psi}$ using Eq. (\ref{potencial}). Finally, we obtain the equation of state $T = T(\tilde{\psi}, \tilde{Q})$, and by solving for $\tilde{\psi}$, we find that
\begin{equation}\label{estado}
\Tilde{\psi}= \frac{32 \pi^3  l_{Pl}^2 \Tilde{Q}_{kk}^2 }{(8\pi l_{Pl}^2\Tilde{Q}_{kk}^2-1)^2}T^2.
\end{equation}
Note that as $\tilde{\psi}\rightarrow\infty$, $8\pi l_{Pl}^2\tilde{Q}_{kk}^2$ becomes unity for a fixed $T$. With the equation of state at hand, we can check if there are critical points. These points are determined by the following conditions:
\begin{equation}\label{crit}
    \bigg(\frac{\partial \Tilde{Q}_{kk}}{\partial\Tilde{\psi}}\bigg)_{T_{crit}}=   \bigg(\frac{\partial^2 \Tilde{Q}_{kk}}{\partial\Tilde{\psi}^2}\bigg)_{T_{crit}}=0.
\end{equation}
However, it is possible to show that these equations cannot admit any critical point for a positive real value of $\tilde{\psi}$. To conclude, the isothermal charge susceptibility (\ref{estab}) for the equation of state (\ref{estado}) is given by
\begin{equation}
    \chi_{T} = -\frac{\pi  T}{2 \Tilde{r}^6+3 \pi  \Tilde{r}^4 T}.
\end{equation}
Immediately, we observe that this quantity is negative, indicating that our solution is electrically unstable. Furthermore,  it can be inferred that our solution does not present critical phenomena, since Eqs. (\ref{crit}) lack solutions and no discontinuities appear in the graph for heat capacity.

\section{Final remarks and perspectives } \label{sec4}
We investigated the impact of the dilaton coupling on Bumblebee excitations in (3+1)-dimensional black hole solutions. By performing a Kaluza-Klein reduction, we obtained a model that combines the Kostelecky-Samuel theory with the dilaton, featuring a positive Liouville potential. In the presence of a static and spherically symmetric black hole, the transverse and longitudinal modes become decoupled from the radial vacuum expectation value (VEV). We then constructed a model that couples the dilaton to these modes in a similar manner. Additionally, the dilaton potential is generated by the massive mode $\beta_0$, while the massless mode of Nambu-Goldstone gives rise to a Maxwell-type field.

We have obtained a charged black hole solution that does not exhibit asymptotic flatness or de Sitter (anti-) behavior.  To ensure the spontaneous breaking of Lorentz symmetry, we derived a constraint that guarantees the positivity of the potential. The solution features an event horizon that is shifted compared to the Schwarzschild horizon due to the presence of LV. Further, the solution is singular at the origin.

To conclude the paper, we conducted an analysis of the thermodynamic stability of the system with microscopic fluctuations. For this analysis, we considered one extra dimension compactified on a circle $S^1$. We discovered that the temperature and entropy associated with the solution are significantly different from those of the Schwarzschild and Reissner-Nordström solutions, illustrating the influence of dilaton dynamics on the solutions. We observed that the temperature decreases as the black hole mass decreases, and the extreme limit ($T=0$) is reached when the KK charge is on the order of the Planck mass. Additionally, we assumed that the entropy follows the area law and demonstrated that the first law of thermodynamics holds for our system. Furthermore, we established a critical limit determined by the constraint (\ref{ine}), indicating a potential phase transition for a system with spontaneous Lorentz symmetry breaking.

Regarding stability, we discovered that the specific heat capacity of our solution is positive, indicating thermodynamic stability. However, we also found a negative isothermal charge susceptibility, indicating electrical instability in our solution. This result suggests a deeper analysis on the relation of the geometry of a static black hole and with spherical symmetry in the propagation of the transverse bumblebee mode.

\appendix
\section{Bumblebee dimension reduction}\label{apen}

In this section, we first discuss how Kaluza-Klein dimensional reduction modifies the equations of the KS (Kostelecky-Samuel) model \cite{kos1}. We start with the action KS in (D+d)-dimensions given by
\begin{equation} \label{action}
    S_{KS} = \int d^{D+d}x \sqrt{-\Tilde{g}}\bigg[\frac{\Tilde{R}}{2 \kappa^2_{D+d}} - \frac{1}{4}B^2_{2} - \frac{\lambda}{2}(B^{M}B_{M}\pm b^2)^2 \bigg], 
\end{equation}
where $\kappa^2_{D+d}=8\pi G_{D+d}$ with $G_{D+d}$ the Newton gravitational constant in (D+d)-dimensions, the $\Tilde{g}$ and the $\Tilde{R}$ are metric determinant and curvature scalar, respectively, in (D+d)-dimensions. We also have a 2-form set to $B_{2}:=\frac{1}{2} B_{MN}dx^{M} \wedge dx^{N}$, where $B_{MN}=\partial_{[M}B_{N]}$ and $B_{M}$  is known as Bumblebee field. The quadratic potential chosen induces the spontaneous Lorentz violation, where $\lambda$ is a mass dimension one positive self-interaction coupling constant, $b^{2}$ is a positive constant with squared mass dimension and the $\pm$ sign meaning if $b_{M}$ is spacelike or timelike. Moreover, the vacuum condition $V=0$ implies the existence of a vacuum expectation value $<B_{M}>=b_{M}$ is the form
\begin{equation} \label{normaa}
g^{MN}b_{M}b_{N} = \mp b^{2}.
\end{equation}

Note that the trade-off between the VEV and the curvature of spacetime is felt from Eq. (\ref{norma}), so that we can determine the components of the background vector $b_{\mu}$ knowing the spacetime we are interested in. Later we will see how a spacelike VEV modifies gravitational objects with spherical and static symmetry, such as black holes.

Now to achieve dimension reduction of the action (\ref{action}), we adopt the following line element
\begin{equation} \label{line}
    ds^{2}_{D+d} = e^{2 \alpha(x)}   ds^{2}_{D} + e^{2 \beta(x)}  ds^{2}_{d},
\end{equation}
where $ ds^{2}_{d}=\delta_{ij}dy^{i} dy^{j} $ is a d-dimensional Euclidean flat space, i.e., the  scalar curvature vanishes and the $\alpha$ and $\beta$  are functions of the coordinate $x$ of subspace described by $ ds^{2}_{D}$.  In order to do the dimensional reduction of the action (\ref{action}) in the direction of the coordinates $x$ and that the compactified theory is also written in Einstein frame, we assume that $\beta = \frac{(2 -D)\alpha}{d}$. Thus, the KS action in D-dimensions in the $x$-coordinate direction is given by
\begin{align} \label{action1a}\nonumber
    &S_{KS} =  \mathcal{V} \int d^{D}x \sqrt{-g}\bigg[\frac{1}{2 \kappa^2_{D+d}}\bigg(R - \frac{1}{2}(\partial\phi)^2 \bigg) - \frac{e^{2\Tilde{a}\phi}}{4}B^2_{2}\\
    &- \frac{\lambda e^{2\Tilde{a}\phi}}{2}(B^{M}B_{M}- b^2)^2 \bigg], 
\end{align}
where  $\mathcal{V}$  is the volume of the subspace described by $ds^{2}_{d}$. Moreover, the $R$ and the $\sqrt{-g}$ are the quantities defined in $ ds^{2}_{D}$. Note that the above action is in  Einstein conformal frame. The scalar field $\phi$  is obtained through the following definition
\begin{equation}
    \alpha^2 =\bigg(\frac{d}{2(D-2)(D+d-2)}\bigg)\phi^2
\end{equation}
and the parameter $\Tilde{a}^2= \frac{d}{2(D-2)(D+d-2)}$ is the coupling constant between the fields and the scalar field, known as dilaton field. Note also that we have not taken into account the vector degrees of freedom in Eq. (\ref{line}). Furthermore, the KK reduction of pure $(D + 1)$-dimensional gravity, i.e., $d=1$,
fixes the coupling constant as follows
\begin{equation}\label{kk}
     \Tilde{a}_{kk}^2= \frac{1}{2(D-2)(D+1)}.
\end{equation}
An interesting case that we will address later on is when $D=4$.

\end{document}